\documentclass[iop,apj,tighten,twocolappendix,numberedappendix]{emulateapj}
\usepackage{apjfonts}

\usepackage{graphicx}
\usepackage{amsmath}
\usepackage{amssymb}
\usepackage{color}
\usepackage{mathtools}
\usepackage{url}

\usepackage[breaklinks,colorlinks,citecolor=blue,linkcolor=red]{hyperref} 
\usepackage[all]{hypcap}

\newcommand\msun{\, \rm M_\odot}

\newcommand\kms{\, \rm km\,s^{-1}}

\begin{document}

\title{Black hole--neutron star mergers are unlikely multi-messenger sources}

\author{Giacomo Fragione\altaffilmark{1,2}}
 \affil{$^1$Center for Interdisciplinary Exploration \& Research in Astrophysics (CIERA), Evanston, IL 60202, USA} 
  \affil{$^2$Department of Physics \& Astronomy, Northwestern University, Evanston, IL 60202, USA}

\begin{abstract}
The promise by the LIGO/Virgo/Kagra (LVK) collaboration to detect black hole--neutron star (BH--NS) mergers via gravitational wave (GW) emission has recently been fulfilled with the detection of GW200105 and GW200115. Mergers of BH--NS binaries are particularly exciting for their multi-messenger potential, since the GW detection can be followed by an electromagnetic (EM) counterpart (kilonova, gamma-ray burst, afterglow) that can reveal important information on the equation of state (EOS) of NSs and the nature of the BH spin. We carry out a statistical study of the binary stars that evolve to form a BH--NS binary and compute the rate of merger events that can be followed by an EM counterpart. We find that $\gtrsim 50\%$ of the mergers can lead to an EM counterpart only in the case BHs are born highly spinning ($\chi_{\rm BH}\gtrsim 0.7$), while this fraction does not exceed about $30\%$ for stiff NS EOSs and a few percent for soft NS EOSs for low-spinning BHs ($\chi_{\rm BH}\lesssim 0.2$), suggesting that a high rate of EM counterparts of BH--NS would provide support for high natal BH spins. However, the possibilities that BHs are born with near-maximal spins and that NS internal structure is described by a stiff EOS are disfavored by current LVK constraints. Considering that these values only represent an upper limit to observe an EM counterpart due to current observational limitations, as in brightness sensitivity and sky localization, BH--NS mergers are unlikely multi-messenger sources.
\end{abstract}

\section{Introduction}
\label{sect:intro}

Mergers of black hole-neutron star (BH--NS) binaries are particularly interesting since they can produce an electromagnetic (EM) counterpart associated with the gravitational (GW) signal at merger. The condition for an EM counterpart to occur is that during the merger phase, which follows the inspiral due to GW emission, the NS does not directly plunge into the BH, but rather is tidally disrupted. For these systems only, a post-merger phase is expected, during which NS matter debris can be ejected or accreted onto the BH producing a luminous event. BH--NS mergers that produce an EM counterpart can provide constraints on the BH spin and accretion process and unique information on the nuclear equation of state (EOS) \citep[e.g.,][]{Pannarale2011,tsang2012,PannaraleBerti2015,FoucartHinderer2018,AscenziDeLillo2019,HindererNissanke2019,ZappaBernuzzi2019,FragioneLoeb2021,TiwariEbersold2021}.

Hundreds of merging BH--NS binaries are expected to be detected in the next few years. The second Gravitational Wave Transient Catalog by the LIGO/Virgo/Kagra (LVK) Collaboration reports tens of BH--BH mergers, two NS--NS mergers, and a candidate BH--NS merger \citep[GW190426;][]{lvc2020cat}. Recently, the first two confirmed BH--NS mergers, GW200105 and GW20115, from the third observational run have been publicly released \citep{AbbottAbbott2021}.

The astrophysical origin of BH--NS mergers remains highly uncertain. The most promising scenario is represented by BH--NS mergers produced as a result of the evolution of field binaries. Within the uncertainties of stellar evolution models, this channel predicts merger rates consistent with the empirical LVK estimate \citep[e.g.,][]{demink2016,gm2018,kruc2018,BroekgaardenBerger2021,ShaoLi2021}. BH--NS binaries do not efficiently form in a dense star cluster, resulting in a merger rate several orders of magnitude smaller than LVK estimated rates \citep[e.g.,][]{clausen2013black,bae2014compact,belczynski2018origin,FragioneBanerjee2020,asedda2020,ye2020}, which can increase under favorable initial conditions \citep{rastello2020}. BH--NS mergers from massive triples as a result of the Lidov-Kozai mechanism have been proposed as an alternative channel, but merger rates consistent with the LVK results are obtained only if very low natal kicks for BHs and NSs are assumed \citep[e.g.,][]{frl2019a,frl2019b}. Finally, BH--NS binaries can be assembled and merge in AGN disks at high rates, but there are still major uncertainties in the models \citep[e.g.,][]{YangGayathri2020,TagawaKocsis2021}. 

The detection of a short gamma-ray burst (GRB) \citep[GW170817A;][]{AbbottAbbottgrb2017}, of an ultraviolet-optical-infrared transient \citep[AT2017gfo;][]{AbbottAbbottmulti2017}, and of an off-axis jet \citep[e.g.,][]{MarguttiBerger2017,TrojaPiro2017} associated with GW170817 \citep{AbbottAbbottnsns2017} has opened the long-awaited era of multi-messenger astronomy for double NS mergers. The question whether BH--NS mergers are typically expected to have EM counterparts remains still to be answered. Many detailed follow-up observations for the candidate and confirmed LVK BH--NS mergers have shown no associated EM counterpart \citep{HosseinzadehCowperthwaite2019,GoldsteinAndreoni2019,CoughlinDietrich2020,ThakurDichiara2020,AnandCoughlin2021,AlexanderSchroeder2021,KilpatrickCoulter2021}. On one hand, the lack of detections could simply be due to the fact that present searches are not sensitive enough (BH--NS mergers can be observed via GW emission at larger distances with respect to binary NSs) and sky localization is not optimal. On the other hand, there is the possibility that EM counterparts are intrinsically missing because the NS typically plunges into the BH, leaving behind no debris to accrete.

In this Letter, we use population synthesis of BH--NS systems from field binaries to study the rate of expected EM counterparts. We adopt different models for natal kicks, efficiencies for common-envelope ejection, BH birth spins, and NS EOSs. We show that a significant fraction of BH--NS mergers is expected to be followed by an EM counterpart only if BHs are born highly-spinning and NSs have stiff EOSs, both of which are currently disfavored by LVK constraints. 

This Letter is organized as follows. In Section~\ref{sect:models}, we describe our models and methods to build a cosmologically-motivated population of BH--NS mergers that can be followed by an EM counterpart. In Section~\ref{sect:res}, we discuss the expected distributions and rates of EM counterparts to BH--NS mergers. Finally, in Section \ref{sect:conc}, we summarize our findings and draw our conclusions.

\section{Method}
\label{sect:models}

\subsection{Binary population synthesis}

We sample the initial mass $m_1$ of the primary from a \citet{kroupa2001} initial mass function,
\begin{equation}
\frac{\mathrm{d}N}{\mathrm{d}m} \propto m^{-2.3},
\label{eqn:massfunc}
\end{equation}
in the mass range $[20\,\msun$--$150\,\msun]$, appropriate for BH progenitors. To determine the initial secondary mass ($m_2$), we adopt a flat mass ratio distribution \citep{sana12,duch2013,Sana2017}. Finally, we extract orbital periods (in days) from
\begin{equation}
f(\log_{10} P) \propto (\log_{10} P)^{-0.55}
\end{equation}
in the range $[0.15$--$5.5]$ \citep{sana12} and assume a thermal distribution for the eccentricity.

The binaries we sample are evolved using the stellar evolution code \textsc{bse} \citep{hurley2000comprehensive,hurley2002evolution}. We use the latest version of \textsc{bse} from \citet{BanerjeeBelczynski2020}, which includes the most up-to-date prescriptions for stellar winds and remnant formation, and produces remnant populations consistent with those from \textsc{StarTrack} \citep{belc2008}.  

We model the distribution of natal velocity kicks (due to recoil from an asymmetric supernova explosion) imparted to compact objects at birth as a Maxwellian distribution with velocity dispersion $\sigma$. In our main model, we assume $\sigma=265$ km s$^{-1}$, consistent with the distribution inferred from proper motions of pulsars by \citet{hobbs2005}. However, the value of $\sigma$ is uncertain. For example, \citet{arz2002} found a bimodal distribution with characteristic velocities of $90\kms$ and $500\kms$ based on the velocities of isolated radio pulsars, while \citet{BeniaminiPiran2016} found evidence for a low-kick population ($\lesssim 30\kms$) and a high-kick population ($\gtrsim 400\kms$) based on observed binary NSs. Therefore, we also run additional models with $\sigma=10$ km s$^{-1}$, $40$ km s$^{-1}$, $150$ km s$^{-1}$. For BHs, we sample natal kicks from the same distribution as for NSs, scaling them down with increasing mass fallback fraction \citep{RepettoDavies2012,Janka2013}. We also self-consistently keep track of the spin-orbit misalignment ($I_{\rm BH-NS}$) produced as a result of natal kicks, by extracting them from \textsc{bse} and computing the tilt of the binary orbit (whenever the orbit remains bound, as outlined in Sect.~2 in \citet{FragioneLoebRasio2021}). We do not model other possible sources of spin-orbit misalignment, as gas torques due to accretion during common-envelope events.

We consider three different models for BH spins. In the first model, we assume that the BH spin is $\chi_{\rm BH}=0.9$, which is consistent with the prescriptions of the \textsc{geneva} stellar evolution code over the mass range relevant for BH--NS mergers \citep{EggenbergerMeynet2008,EkstromGeorgy2012}. In the second model, we assume $\chi_{\rm BH}=0.1$, consistent with the results of the \textsc{mesa} stellar evolution code \citep{PaxtonBildsten2011,PaxtonMarchant2015}. Indeed, \textsc{mesa} includes a treatment for the magnetic field that makes the outwards angular momentum transport from the core much more efficient by forming a Tayler-Spruit magnetic dynamo, resulting in a small BH spin. Recent detailed studies of massive stars have suggested that the Tayler-Spruit magnetic dynamo can essentially extract all of the angular momentum, birthing BHs with extremely low spins \citep{FullerMa2019}. Therefore, we also consider a model where the initial spin of BHs is assumed to vanish. For full details see \citet{baner2019bse} and \citet{Banerjee2021}.

\begin{figure} 
\centering
\includegraphics[scale=0.595]{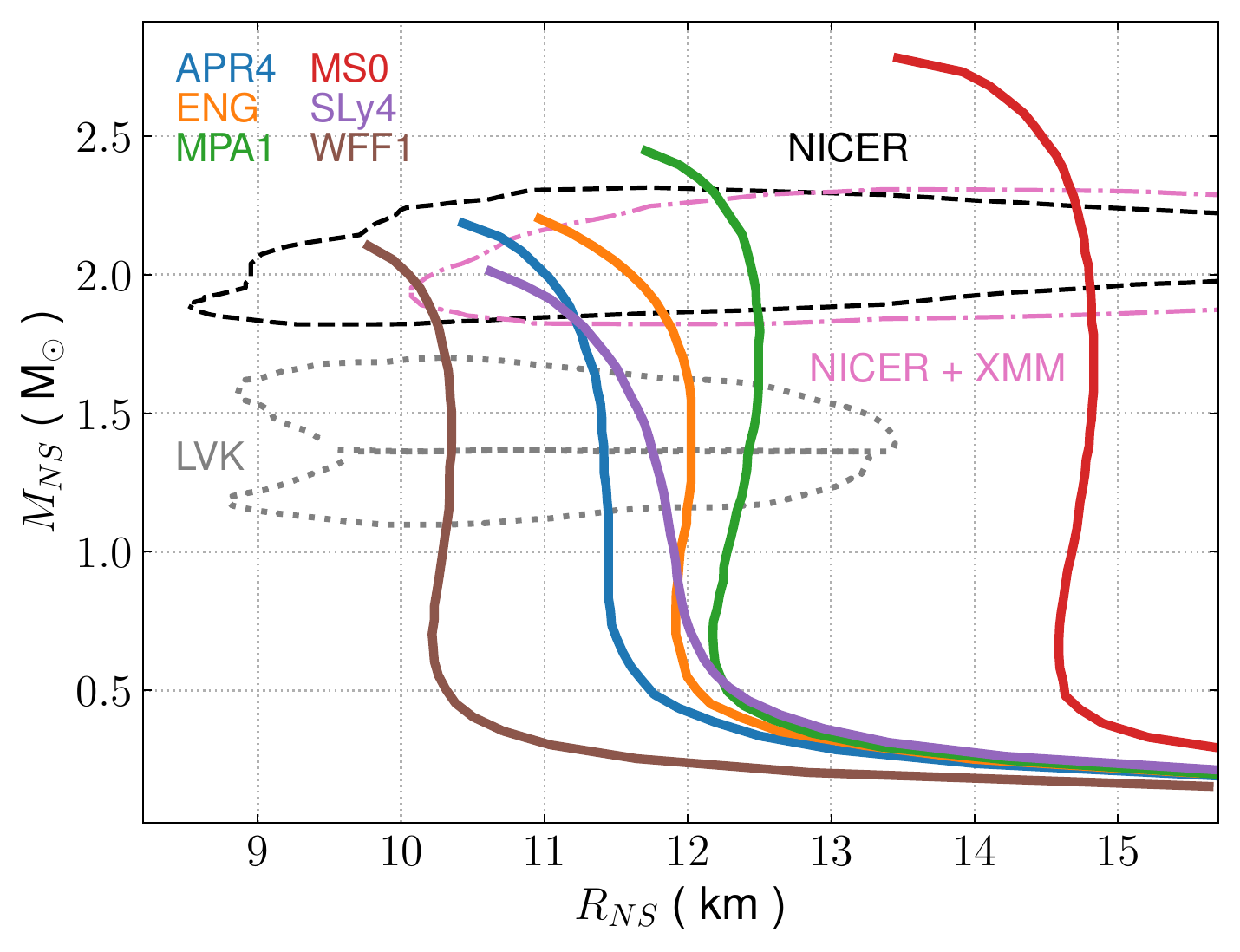}
\caption{Observational constraints on the mass-radius relationship compared with NS equilibrium sequences. The dotted-gray line represents constraints ($95\%$ confidence region) for low-mass and high-mass NSs from LVK using data from the NS-NS merger GW170817, while the black-dashed line and the pink-dot-dashed line represent constraints ($95\%$ confidence region) from NICER and NICER+XMM, respectively, using PSR J0740+6620.}
\label{fig:eos}
\end{figure}

Finally, we consider stellar evolution with $6$ different assumed metallicities, namely $Z=0.0002$, $0.001$, $0.002$, $0.005$, $0.01$, $0.02$, and three different values of the common-envelope energy efficiency parameter, $\alpha_{\rm CM}=1$, $3$, $5$ \citep{hurley2002evolution}. 
\subsection{Electromagnetic counterpart}

To compute whether a BH--NS merger produces an EM signature, we compute the remnant baryon mass ($M_{\rm rem}$) outside the BH after merger. We assume that if $M_{\rm rem}>0$ a disk is formed and there is EM emission, otherwise the NS plunges directly into the BH leaving no signatures other than the GW inspiral \citep[e.g.,][]{Foucart2012,FoucartDeaton2013}. To estimate the remnant disk mass (in units of the initial mass of the NS), we use
\begin{equation}
\hat{M}_{\rm rem}=\left[\max\left(\alpha \frac{1-2C_{\rm NS}}{\eta^{1/3}}-\beta\hat{R}_{\rm ISCO}\frac{C_{\rm NS}}{\eta}+\gamma,0\right)\right]^\delta\,,
\label{eqn:mrem}
\end{equation}
where $(\alpha,\beta,\gamma,\delta)=(0.406,0.139,0.255,1.761)$ are fitting parameters to numerical relativity simulations \citep{Foucart2012,FoucartDeaton2013,FoucartHinderer2018}, $C_{\rm NS}=GM_{\rm NS}/(R_{\rm NS}c^2)$ is the NS compaction, which depends on the EOS, $\eta=Q/(1+Q)^2$, with $Q=M_{\rm BH}/M_{\rm NS}$, is the symmetric mass ratio, and (in units $G=c=1$),
\begin{equation}
\hat{R}_{\rm ISCO}=\frac{R_{\rm ISCO}}{M_{\rm BH}}=3+Z_2-{\rm sgn}(\chi_{\rm BH,\parallel})\sqrt{(3-Z_1)(3+Z_1+3Z_2)}\,,
\end{equation}
is the innermost stable circular orbit (ISCO) radius \citep{BardeenPress1972}, with 
\begin{equation}
Z_1=1+(1-\chi_{\rm BH,\parallel}^2)^{1/3}[(1+\chi_{\rm BH,\parallel})^{1/3}+(1-\chi_{\rm BH,\parallel})^{1/3}]
\end{equation}
\begin{equation}
Z_2=\sqrt{3\chi_{\rm BH,\parallel}^2+Z_1^2}\,.
\end{equation}
In the previous equations, $\chi_{\rm BH,\parallel}$ is the aligned component of the BH spin ($\chi_{\rm BH,\parallel}=\chi_{\rm BH}\cos I_{\rm BH-NS}$) with respect to the orbital angular momentum.

For the NS EOS, we consider six different equilibrium sequences, namely APR4 \citep{AkmalPandharipande1998}, ENG \citep{EngvikOsnes1996}, MPA1 \citep{MutherPrakash1987}, MS0 \citep{MullerSerot1996}, SLy4 \citep{ChabanatBonche1998}, and WFF1 \citep{WiringaFiks1988}. We show these NS equilibium sequences in Figure~\ref{fig:eos} along with current observational constraints from LVK \citep{AbbottAbbotteos2018} and NICER(+ XMM) \citep{MillerLamb2021}. Note that, while NICER constraints allow NSs with large radii (stiff EOSs), this portion of the parameter space is excluded by LVK constraints at $95\%$ confidence, favoring softer EOSs.

\subsection{Distribution of BH--NS mergers with electromagnetic counterpart}

To place our BH--NS population in a cosmological context, we assign to each binary a formation time $t_{\rm form}$ by sampling the formation redshift $z_{\rm form}$ from the cosmic star formation history of \citet{MadauDickinson2014}
\begin{equation}
\Psi(z)=0.01 \frac{(1+z)^{2.6}} {1.0 + [(1.0 + z) / 3.2]^{6.2}}\,{\rm M}_\sun\,{\rm yr}^{-1}\,{\rm Mpc}^{-3}\,.
\label{eqn:madau}
\end{equation}
Then, we convolve the delay time $t_{\rm delay}$ of the BH--NS mergers with the distribution of formation times and discard the binaries that merge later than the present day. Each BH--NS that is not discarded is then assigned a weight that accounts for the cosmic distribution of metallicity, which we assume is described by a log-normal distribution $\Pi(z,Z)$, with mean given by \citep{MadauFragos2017}
\begin{equation}
\log \langle Z/{\rm Z}_\odot \rangle = 0.153 - 0.074 z^{1.34}
\end{equation}
and a standard deviation of 0.5 dex \citep{DvorkinSilk2015}. This weighting procedure provides us with the underlying astrophysical distribution of sources at a given redshift interval per comoving volume.

\begin{figure} 
\centering
\includegraphics[scale=0.575]{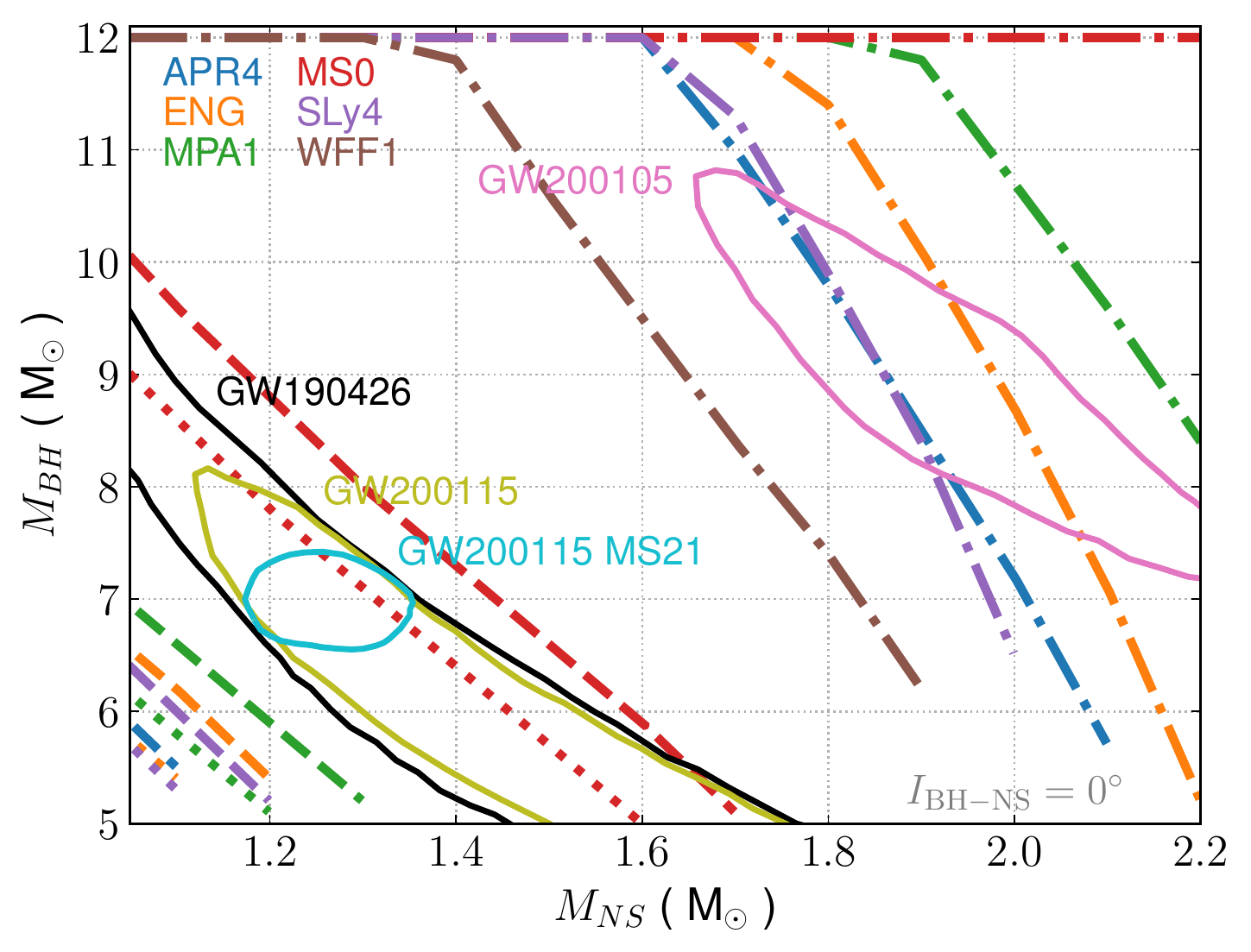}
\includegraphics[scale=0.575]{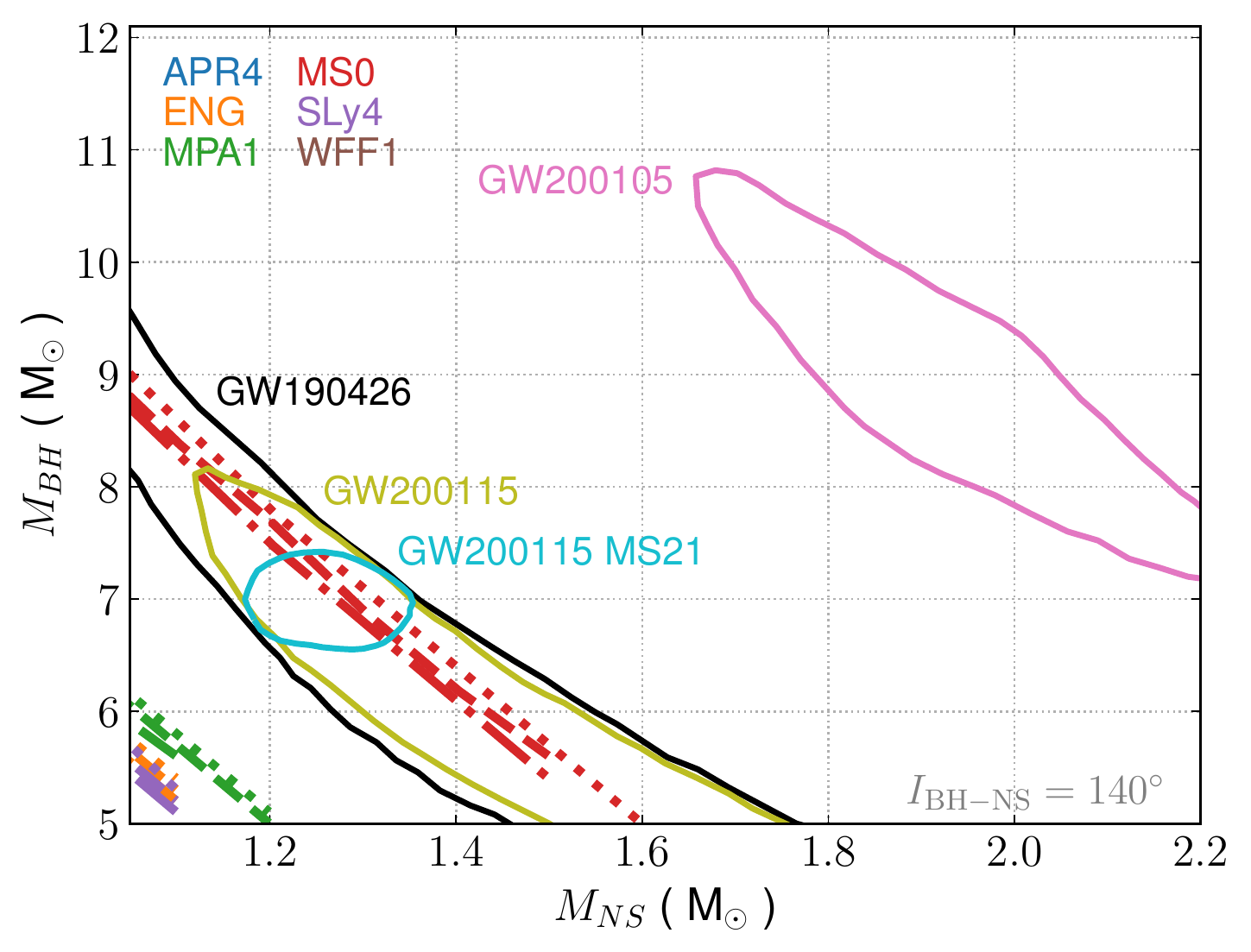}
\caption{The region below each line covers the portion of the mass parameter space that allows the tidal disruption of a NS by a BH for different EOSs. Different line styles represent different BH spin models: $\chi_{\rm BH}=0.9$ (dot-dashed), $\chi_{\rm BH}=0.1$ (dashed), $\chi_{\rm BH}=0.0$ (dotted). Top panel: $I_{\rm BH--NS}=0^\circ$; bottom panel: $I_{\rm BH--NS}=140^\circ$. Solid lines show the 90\% credible interval for the LVK candidate event GW190426, the two confirmed LVK BH--NS mergers GW201005 and GW20015, and the re-analysis by \citet{MandelSmith2021} of GW200115 (GW200115 M21).}
\label{fig:example}
\end{figure}

\begin{figure*} 
\centering
\includegraphics[scale=0.55]{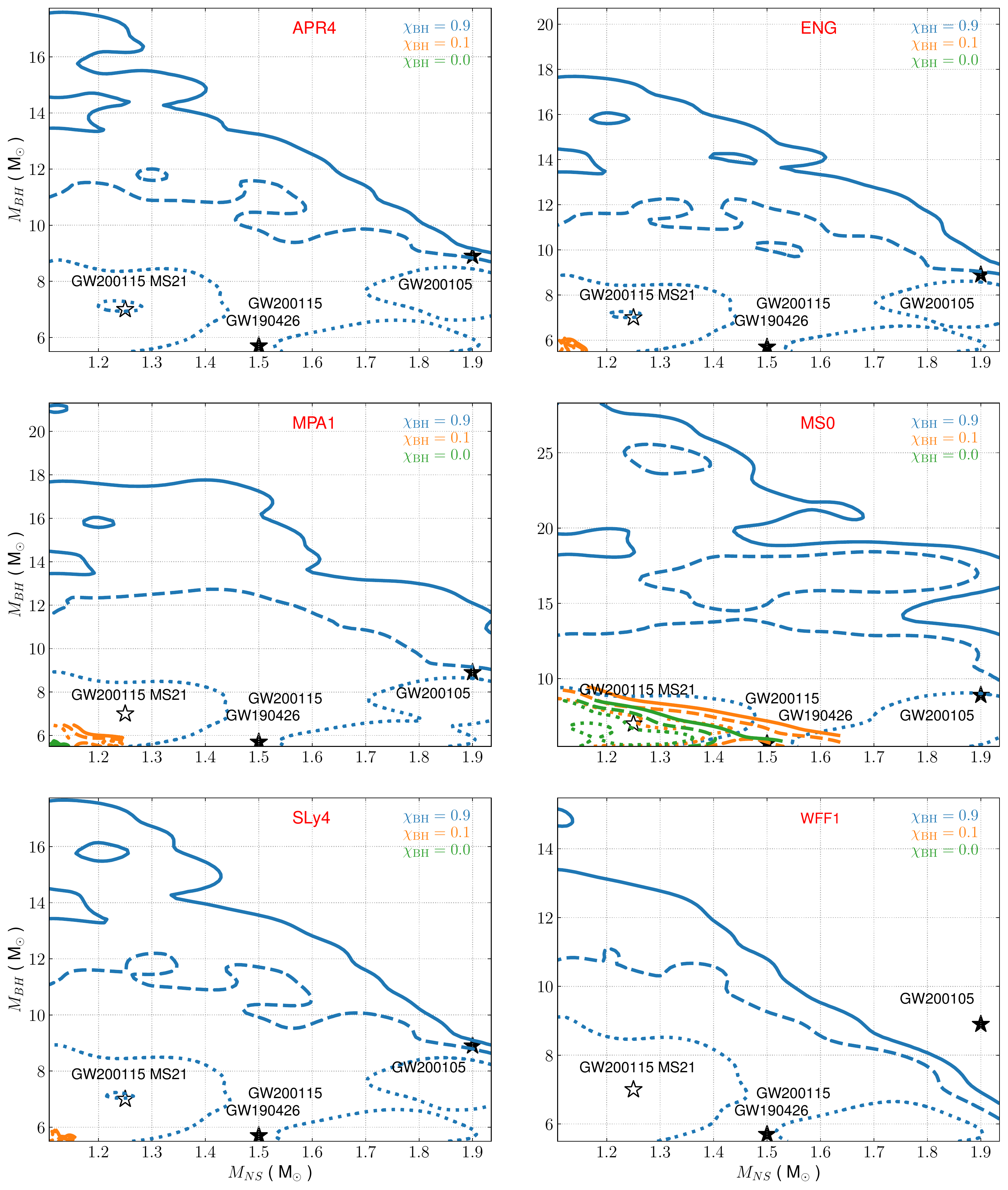}
\caption{Component masses of BH--NS mergers that produce an EM counterpart ($\sigma=260\kms$ and $\alpha_{\rm CM}=1$). Contour lines represent $68\%$ (dotted), $95\%$ (dashed), $99.7\%$ (solid) of the reconstructed astrophysical population. Star symbols show the median mass for the candidate event GW190426, two confirmed LVK BH--NS mergers GW201005 and GW20015, and the re-analysis by \citet{MandelSmith2021} of GW200115 (GW200115 M21; void symbol). Different colors represent different assumptions on the BH spin, while different panels show results for different EOSs (in red).}
\label{fig:m1m2}
\end{figure*}

\begin{figure*}
\centering
\includegraphics[scale=0.55]{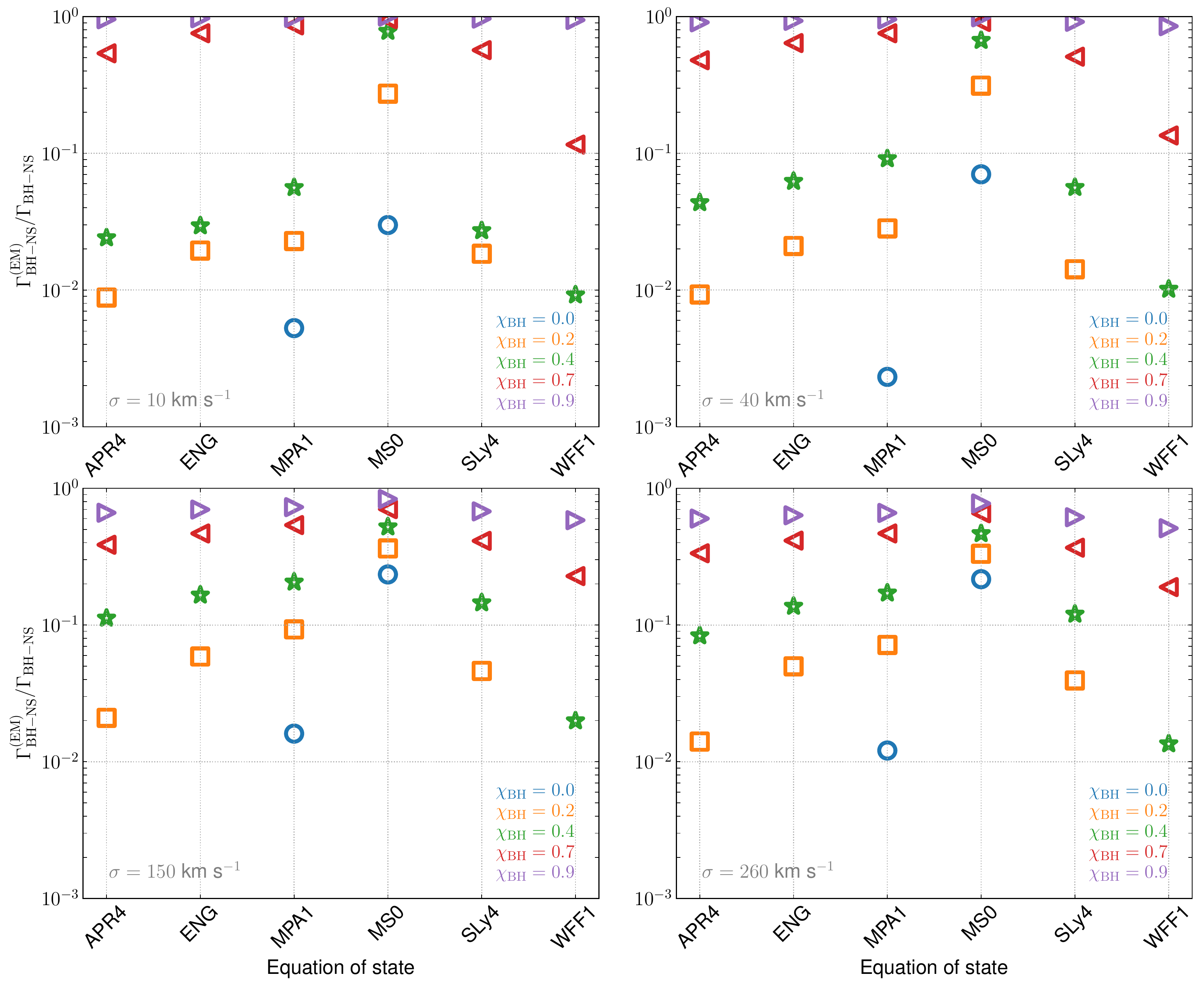}
\caption{Fractional rate in the local Universe of BH--NS mergers that have an EM counterpart as a function of the NS EOS ($\alpha_{\rm CM}=1$). Different panels represent different values of $\sigma$, while different symbols represent different spin models: $\chi_{\rm BH}=0.0$ (blue circles), $\chi_{\rm BH}=0.2$ (orange squares), $\chi_{\rm BH}=0.4$ (green stars), $\chi_{\rm BH}=0.7$ (red triangles), $\chi_{\rm BH}=0.9$ (purple triangles).}
\label{fig:rates1}
\end{figure*}

\subsection{Merger rates of BH--NS mergers with electromagnetic counterpart}

For a specific EOS, we compute the rates of BH--NS mergers with electromagnetic counterpart as
\begin{eqnarray}
\Gamma^{\rm (EM)}_{\rm BH-NS}(z) &=& f_{\rm b} f_{\rm IMF} \frac{d}{dt_{\rm lb}(z)} \int^{z_{\max}}_{z} \Psi(\zeta) \frac{dt_{\rm lb}(\zeta)}{d\zeta} d\zeta \nonumber\\
& \times & \int^{Z_{\max}(\zeta)}_{Z_{\min}(\zeta)} \Phi^{\rm (EM)}(z, Z) \Pi(\zeta, Z) dZ\,,
\label{eqn:ratez}
\end{eqnarray}
where $f_{\rm b}=0.5$ is the fraction of stars in binaries \citep[e.g.,][]{ragh10,duch2013,Sana2017}, $f_{\rm IMF}=0.115$ is a correction factor that accounts for our truncation of the primary mass distribution $\ge 20\msun$ (assuming a \citet{kroupa2001} initial mass function), $t_{\rm lb}$ is the look-back time at redshift $z$ \footnote{For our calculations, we assume the cosmological parameters from Planck 2015 \citep{PlanckCollaborationAde2016}.}. In Eq.~\ref{eqn:ratez}, $\Phi$ is the merger efficiency at a given metallicity of binaries that produce an EM counterpart
\begin{equation}
\Phi^{\rm (EM)}(z, Z) = S^{\rm (EOS)}_{\rm EM}(z,Z) \frac{N_{\rm merger} (z, Z)}{M_{\rm tot} \, (Z)} \, ,
\end{equation}
where $M_{\rm tot} (Z)$ is the total simulated mass at metallicity $Z$ and $N_{\rm merger} (z, Z)$ the total number of BH--NS mergers at redshift $z$ originating from progenitors at metallicity $Z$, and
\begin{equation}
S^{\rm (EOS)}_{\rm EM}(z,Z) = \frac{N^{(EM)}_{\rm merger} (z, Z)}{N_{\rm merger} (z, Z)}
\end{equation}
is the fraction of merging systems that have an EM counterpart ($\hat{M}_{\rm rem}>0$), assuming a given EOS. To compute BH--NS merger rates (considering both systems with and without EM counterparts), $\Gamma_{\rm BH-NS}(z)$, we simply impose $S^{\rm (EOS)}_{\rm EM}(z,Z)= 1$.

\section{Results}
\label{sect:res}

For a given EOS, whether a NS plunges directly onto a BH or is tidally disrupted producing an EM counterpart depends crucially on the BH spin and its orientation with respect to the orbital angular momentum. In Figure~\ref{fig:example}, the region below each line covers the portion of the mass ($M_{\rm NS}$--$M_{\rm BH}$) parameter space that allows the tidal disruption of a NS by a BH for each combination of NS EOSs and BH spin models considered in this work. We also report the 90\% credible interval for the LVK candidate event GW190426, the two confirmed LVK BH--NS mergers GW201005 and GW20015, and the re-analysis by \citet{MandelSmith2021} of GW200115 (GW200115 M21). Note that, for the BH masses of interest here, $\chi_{\rm BH}=0.9$ and $\chi_{\rm BH}=0.1$ correspond to the \textsc{geneva} model and the \textsc{mesa} model, respectively, while $\chi_{\rm BH}=0.0$ to the model of \citet{FullerMa2019} \citep[see Figure~3 in][]{Banerjee2021}. In the top panel, we show the case $I_{\rm BH--NS}=0^\circ$, that is the BH spin is aligned to the orbital angular momentum. Under this assumption, a BH of $\sim 10\msun$ could lead to a tidal disruption of a NS of a typical mass of $1.3\msun$ for even the softest EOSs here considered. Moreover, we find that the stiffer the NS EOS is and the higher the BH spin is, the larger is the parameter space that leads to the tidal disruption of the NS, eventually producing an EM counterpart. We find that LVK events could have had (within mass uncertainty limits) an EM counterpart in the case of highly-spinning BHs. However, current constraints on the effective spin of these events suggest a very low value for the BH spin, rendering the likelihood of an EM counterpart small. The previous trend is highly affected by the orientation of the BH spin with respect to the orbital angular momentum. As an illustrative case, we plot the critical lines in the case $I_{\rm BH--NS}=140^\circ$ in the bottom panel, which is representative of the median misalignment of the BH spin in GW200115 \citep{AbbottAbbott2021}, even though recent analysis by \citet{MandelSmith2021} cast doubt on this result. In this case, the region of the mass parameter space that allows the tidal disruption of a NS shrinks significantly and does not critically depend on the magnitude of the BH spin. In this case, APR4 and WFF1, the two softest EOS here considered, always lead to a direct plunge, while the other EOSs allow the tidal disruption of a NS only for BH masses $\lesssim 6\msun$. Only MS0, the stiffest EOS we consider, would predict an EM counterpart for GW190426 and GW200115 (within mass uncertainty limits), regardless of the BH spin. Note that, however, the possibility that all BHs are born with near-maximal spins is becoming disfavored \citep{AbbottAbbottpop2021} and very stiff EOSs are excluded at 95\% confidence by LVK constraints on the NS EOS \citep{AbbottAbbotteos2018}. Moreover, their kilonova brightness would be too faint to be detected for present follow-up search campaigns, justifying the lack of detections so far \cite[e.g.,][]{ZhuWu2021}.

To understand whether BH--NS mergers are typically expected to be multi-messenger sources, we illustrate in Figure~\ref{fig:m1m2} the contours ($68\%$, $95\%$, $99.7\%$) of component masses of BH--NS mergers that produce an EM counterpart in our cosmologically-motivated model for different EOSs and BH spin models. In this case, we fix $\sigma=260\kms$ and $\alpha_{\rm CM}=1$ in our population synthesis models; other combinations of $\sigma$ and $\alpha_{\rm CM}$ lead to qualitatively similar results. Note that we account for the cosmological distribution of inclination angles, $I_{\rm BH-NS}$, \citep[see][]{FragioneLoebRasio2021}, which is critical to determine whether a NS directly plunges onto or is tidally disrupted by the BH. We also report in Figure~\ref{fig:m1m2} the median mass for the LVK candidate event GW190426, the two confirmed LVK BH--NS mergers GW201005 and GW20015, and the re-analysis by \citet{MandelSmith2021} of GW200115. Using similar considerations used in the previous plot, we find that a considerable fraction of our cosmologically-motivated population of BH--NS mergers can eventually produce an EM counterpart only in the case BHs are born highly spinning. Under this assumption, current LVK systems would lie in the $95\%$-density region, with the exception of GW201005 in the case we consider WFF1, our softest EOS. If BHs are born slowly spinning, only MS0, our stiffest EOS, allows the tidal disruption of NSs, and the eventual EM counterpart, in a non-negligible fraction of the mass parameter space.

Finally, we compute the fractional rate in the local Universe of BH--NS systems that could lead to an EM counterpart with respect to the total rate of BH--NS mergers, for different values of $\chi_{\rm BH}$. We show our results in Figure~\ref{fig:rates1}, for different values of $\sigma$ and assuming $\alpha_{\rm CM}=1$. We report our results for different values of $\alpha_{\rm CM}$ in Appendix \ref{sect:app}. We find that $\gtrsim 50\%$ of the mergers can lead to an EM counterpart only in the case BHs are born highly spinning ($\chi_{\rm BH}\gtrsim 0.7$). In the case BH are born with small spins ($\chi_{\rm BH}\lesssim 0.2$), the fraction of systems that could have an EM counterpart does not exceed about $30\%$ for MS0, our stiffest EOS. These trends appear to be qualitatively the same for different values of $\alpha_{\rm CM}$ and $\sigma$, since high natal kicks imply larger spin-orbit misalignment \citep[see Fig.~2 in][]{FragioneLoebRasio2021}, but, at the same time, they may disrupt the binary system, preventing the BH-NS binary to eventually merge, and affect the shape of the delay-time distribution (time between binary formation and merger). As the EOS of NSs gets better constrained, our results predict that a high rate of detected EM counterparts to BH--NS mergers would support the case that BHs are preferentially born with high spins. The possibility that all BHs are born with near-maximal spins is becoming disfavored \citep{AbbottAbbottpop2021} and very stiff EOSs are excluded at 95\% confidence by LVK constraints on the NS EOS \citep{AbbottAbbotteos2018}. Under these assumptions, we find that the fraction of BH--NS mergers that can possibly have an EM counterpart does not exceed about $10\%$. In any case, the future measured EM counterpart fraction could be translated sensitively into constraints on the BH spin using the results of our Figure~\ref{fig:rates1}. However, note that these values only represent an upper limit of observable EM counterparts, since, for example, they might be too faint to observe their kilonova and sky localization could not be optimal for a possible follow-up observation \cite[e.g.,][]{ZhuWu2021}.

We finally note that in our models we have taken into account only the spin-orbit misalignment produced as a result of natal kicks and have not modeled other possible sources of spin-orbit misalignments, as gas torques due to accretion during common-envelope events. The detailed prescriptions of core-collapse physics may also play a role \citep{Roman-GarzaBavera2021}. Also, we note that there might be other types of EM counterparts associated with BH--NS mergers, as shock-powered radio precursors \citep{SridharZrake2021} and resonant shattering flares \citep{NeillTsang2021}.

\section{Conclusions}
\label{sect:conc}

BH--NS mergers are very interesting since they could provide crucial information on their origin, the nature of the BH spin, and the NS internal structure. However, this relies on the fact that BH--NS mergers are followed by an EM counterpart, which happens whenever the NS does not directly plunge onto the BH.

We have carried out a broad statistical study of the field binary stars that form merging BH--NS binaries and have evaluated the fraction that can eventually be associated with an EM counterpart. We have considered different NS EOSs and BH spin models, and we have taken into account the uncertainties on the natal-kick magnitudes and efficiency of common-envelope ejection for compact objects. We have found that $\gtrsim 50\%$ of BH--NS mergers can lead to an EM counterpart only in the case BHs are born highly spinning ($\chi_{\rm BH}\gtrsim 0.7$), otherwise ($\chi_{\rm BH}\lesssim 0.2$) this fraction does not exceed a few percent for soft EOSs and $30\%$ for stiff EOSs. However, current LVK constraints tend to disfavor the scenario where all BHs are born with near-maximal spins \citep{AbbottAbbottpop2021} and exclude very stiff EOSs at 95\% confidence \citep{AbbottAbbotteos2018}. Considering that these values only represent an upper limit due to current limitations of EM follow-up observations to GW detections, we conclude that BH--NS mergers are unlikely multi-messenger sources.

\section*{Acknowledgements}

We thank the anonymous referee for constructive comments and suggestions that have helped to improve the manuscript. G.F. is grateful to Sambaran Banerjee for useful discussions on stellar evolution and for updating \textsc{bse}. G.F.\ acknowledge support from NASA Grant 80NSSC21K1722.

\appendix

\section{Fractional rates for other common-envelope energy efficiencies}
\label{sect:app}

We report here the fractional rate in the local Universe of BH--NS systems that could lead to an EM counterpart with respect to the total rate of BH--NS mergers (similar to Figure~\ref{fig:rates1}), when assuming $\alpha_{\rm CM}=3$ and $\alpha_{\rm CM}=5$ for the common-envelope energy efficiency in our binary simulations.

\begin{figure*} 
\centering
\includegraphics[scale=0.55]{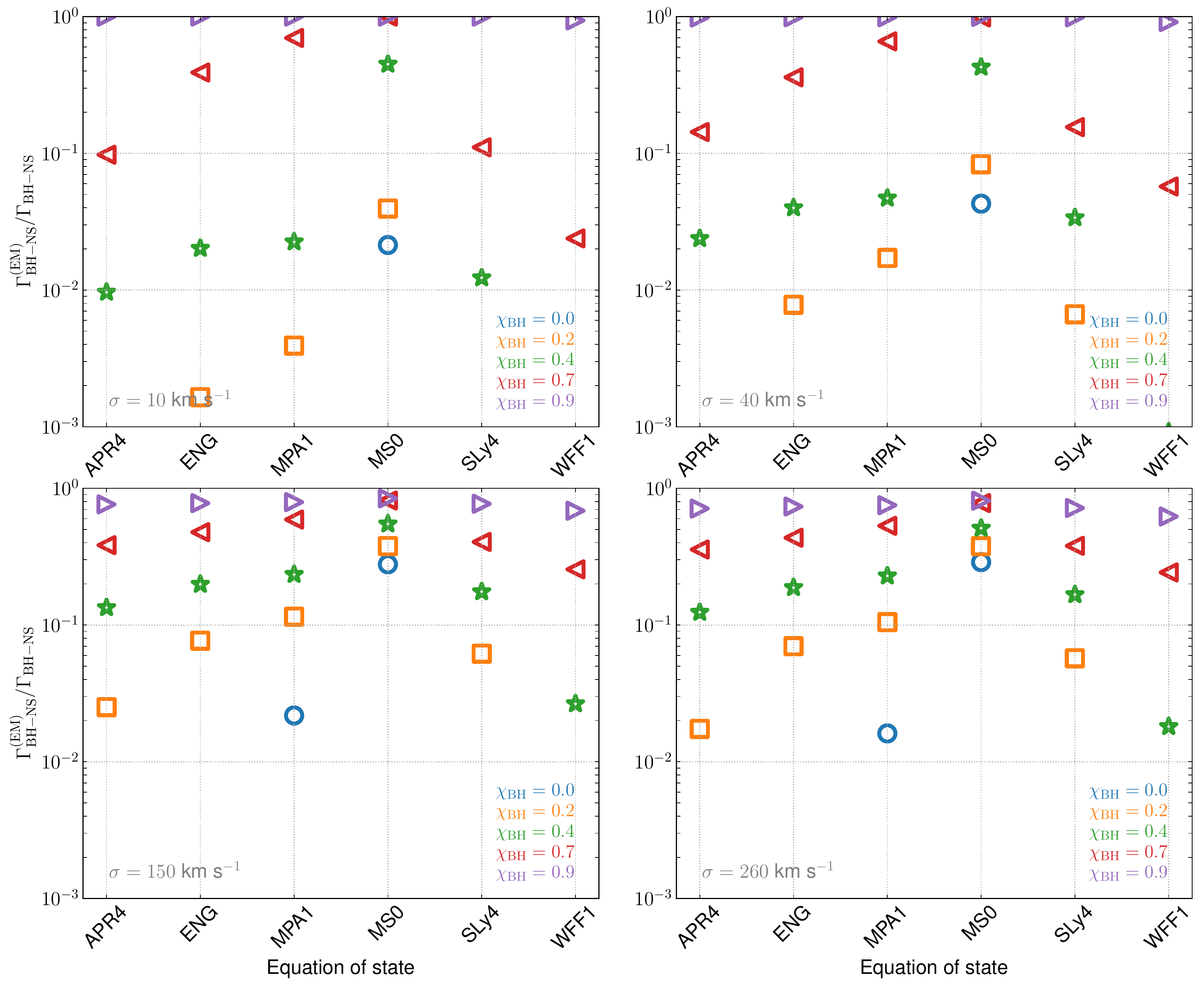}
\caption{Same as Figure~\ref{fig:rates1}, but for $\alpha_{\rm CM}=3$.}
\label{fig:rates3}
\end{figure*}

\begin{figure*} 
\centering
\includegraphics[scale=0.55]{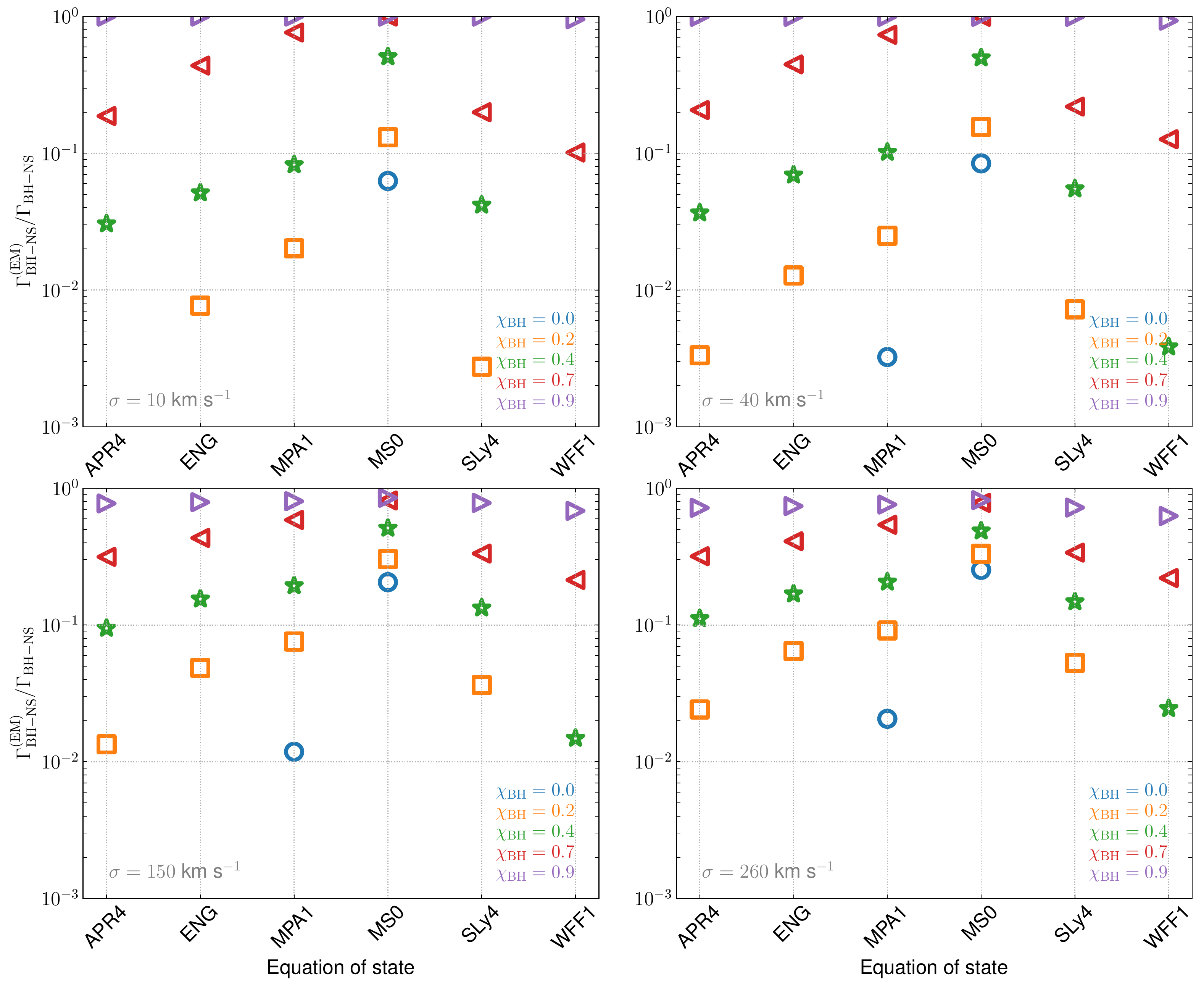}
\caption{Same as Figure~\ref{fig:rates1}, but for $\alpha_{\rm CM}=5$.}
\label{fig:rates5}
\end{figure*}

\bibliographystyle{yahapj}
\bibliography{refs}

\end{document}